# PHYSICS AT A $\mu^+\mu^-$ COLLIDER[a]

J. F. GUNION

*Department of Physics, University of California at Davis, Davis, CA, 95616, USA*

A brief overview of the physics capabilities of a $\mu^+\mu^-$ collider is given, with particular focus on special Higgs sector opportunities.

Although design and development are at a very early stage compared to the next linear $e^+e^-$ collider (NLC), and feasibility is far from proven, there is now considerable interest in the possibility of constructing a $\mu^+\mu^-$ collider;[1] its promise for physics was clear from the beginning.[2] Two specific muon collider schemes are under consideration. A lower energy machine, the First Muon Collider (FMC), could have center-of-mass energy ($\sqrt{s}$) around 0.5 TeV with a luminosity of order $2\times 10^{33}$ cm$^{-2}$ s$^{-1}$ for unpolarized beams.[3] Not only would the FMC be able to accomplish everything that the NLC could (for the same luminosity), but also the FMC would be extremely valuable for discovery and precision studies of Higgs bosons produced directly in the $s$-channel.[4] A high energy next muon collider[3] (NMC) with 4 TeV c.m. energy and luminosity of order $10^{35}$ cm$^{-2}$ s$^{-1}$ would have an energy reach appropriate for pair production of supersymmetric particles and of SUSY Higgs bosons $A^0 H^0$ up to very large masses, and for the study of a strongly interacting $WW$ sector. Here, we give a brief summary of $s$-channel Higgs boson physics at the FMC, followed by an overview of physics at the NMC.[a]

## 1 s-Channel Higgs Physics

For $s$-channel studies of narrow resonances, the energy resolution is crucial. The rms error $\sigma$ in $\sqrt{s}$ is given by $\sigma = (0.04\text{ GeV})\left(\frac{R}{0.06\%}\right)\left(\frac{\sqrt{s}}{100\text{ GeV}}\right)$ where $R$ values of $0.04 - 0.08\%$ are most natural, with $R = 0.01\%$ achievable.[5] A critical issue is how this resolution compares to the calculated total widths of Higgs bosons. For $R \lesssim 0.06\%$, $\sigma$ can be smaller than the Higgs widths in many cases; and for $R \lesssim 0.01\%$ the energy resolution becomes comparable to even the very narrow width of an intermediate-mass SM $\phi^0$. An $s$-channel Higgs resonance could be found by scanning in $\sqrt{s}$ using steps of size $\sim \sigma$; its mass would be simultaneously determined with roughly this same accuracy in the initial scan. For sufficiently small $\sigma$, the Breit-Wigner resonance line-shape would be revealed and the Higgs width could be deduced.

However, the optimal strategy for SM Higgs ($\phi^0$) *discovery* at a lepton collider is to use the $\mu^+\mu^- \to Z\phi^0$ mode (or $e^+e^- \to Z\phi^0$) because no energy scan is needed. Studies of $e^+e^-$ collider capabilities indicate that the SM Higgs can be discovered if $m_{\phi^0} < 0.7\sqrt{s}$. If $m_{\phi^0} \lesssim 140\text{ GeV}$, its mass will be determined to within[6] $\Delta m_{\phi^0} \lesssim 0.4\text{ GeV}\left(\frac{10\text{ fb}^{-1}}{L}\right)^{\frac{1}{2}}$, yielding, *e.g.*, $\pm 180$ MeV for $L = 50$ fb$^{-1}$.[6] At the LHC the $\phi^0 \to \gamma\gamma$ mode is deemed viable for $80 \lesssim m_{\phi^0} \lesssim 150$ GeV, with a better than 1% mass resolution.[7] Once the $\phi^0$ signal is found, precision determination of its mass and width become the paramount goals, for which $s$-channel resonance production at a $\mu^+\mu^-$ collider is uniquely suited.

For $m_{\phi^0} < 2m_W$ the dominant $\phi^0$-decay channels are $b\bar{b}$, $WW^\star$, and $ZZ^\star$, where the star denotes a virtual weak boson. The light quark backgrounds to the $b\bar{b}$ signal can be rejected by $b$-tagging. For the $WW^\star$ and $ZZ^\star$ channels we employ only the mixed leptonic/hadronic modes and the visible purely-leptonic $ZZ^\star$ modes, taking into account the major electroweak QCD backgrounds. For all channels we assume a general signal and background identification efficiency of $\epsilon = 50\%$, after selected acceptance cuts. In the case of the $b\bar{b}$ channel, this is to include the efficiency for tagging at least one $b$. Values of $\epsilon \overline{\sigma} BR(X)$ at $\sqrt{s} = m_{\phi^0}$ for $X = b\bar{b}$, $WW^\star$ and $ZZ^\star$ are presented in Fig. 1 versus $m_{\phi^0}$ for a resolution $R = 0.06\%$. (Here $\overline{\sigma}$ denotes a cross section after convolution with the Gaussian energy spectrum.) The background level ($B$) is essentially independent of $R$, while the signal rate ($S$) depends strongly on $R$.

The luminosity required to achieve $n_\sigma = S/\sqrt{B} = 5$ in the $b\bar{b}$, $WW^\star$ and $ZZ^\star$ channels is also shown in Fig. 1 — results for $R = 0.01\%$, $0.06\%$ and $1\%$ as a function of $m_{\phi^0}$ are illustrated. For $R = 0.06\%$, $L = 1$ fb$^{-1}$ would yield a detectable $s$-channel Higgs signal for all $m_{\phi^0}$ values between the current LEP I limit and $2m_W$ except in the region of the $Z$ peak; a luminosity $L \sim 10$ fb$^{-1}$ at $\sqrt{s} = m_{\phi^0}$ is needed for $85 \lesssim m_{\phi^0} \lesssim 100\text{ GeV}$. For $R = 0.01\%$, $n_\sigma = 5$ signals are achieved with only about $1/25$ of the luminosity required for $R = 0.06\%$, implying that a search by scanning would be most efficient for the smallest possible $R$. If the Higgs resonance is broad, using small $R$ is not harmful since the data from a fine scan can be rebinned to test for its presence.

Once the Higgs is observed, the highest priority will be to determine its precise mass and width. This can be accomplished by scanning across the Higgs peak. The luminosity required for this is strongly dependent upon $R$ (*i.e.* $\sigma$) and the width itself. For a SM Higgs the width in the intermediate mass range is generally smaller than $\sigma$; e.g. for $m_{\phi^0} = 120\,\text{GeV}$ the width is $\Gamma_{\phi^0} \sim 0.004\,\text{GeV}$. A set of carefully chosen measurements is required. The minimal set is three measurements separated in $\sqrt{s}$ by $2\sigma$; the first would be taken at $\sqrt{s}$ equal to the current best central value of the mass (from the initial detection scan). The second and third would be at $\sqrt{s}$ values $2\sigma$ below and $2\sigma$ above the first, with about 2.5 times the integrated luminosity expended on the first. In Fig. 2 we plot the total combined luminosity required for a $\delta\Gamma/\Gamma = 1/3$ measurement of the width in the $b\overline{b}$ channel as a function of $m_{\phi^0}$. For given $R$, luminosity requirements vary by up to 50%, depending upon luck in placement of the first scan point, as quantified by the ratio $|\sqrt{s} - m_{\phi^0}|/\sigma$. The excellent $R = 0.01\%$ resolution would be needed to be certain of being able to measure the total width if $m_{\phi^0} \sim m_Z$. Note that the Higgs mass is also determined to the accuracy of $\sim \delta\Gamma$.

---

[a] To appear in the Proceedings of the International Europhysics Conference on High Energy Physics, Brussels, 1995.

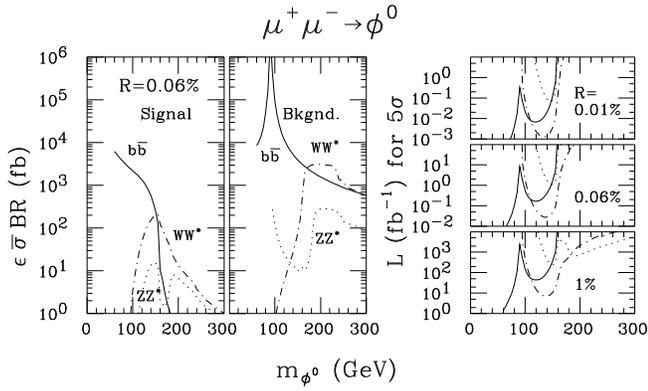

Figure 1: The (a) $\phi^0$ signal and (b) background cross sections, $\epsilon \overline{\sigma} BR(X)$, for $X = b\overline{b}$, and useful $WW^\star$ and $ZZ^\star$ final states (including a channel-isolation efficiency of $\epsilon = 0.5$) versus $m_{\phi^0}$ for SM Higgs $s$-channel production with resolution $R = 0.06\%$. Also shown: (c) the luminosity required for $S/\sqrt{B} = 5$ in the three channels as a function of $m_{\phi^0}$ for $R = 0.01\%$, 0.06% and 1%.

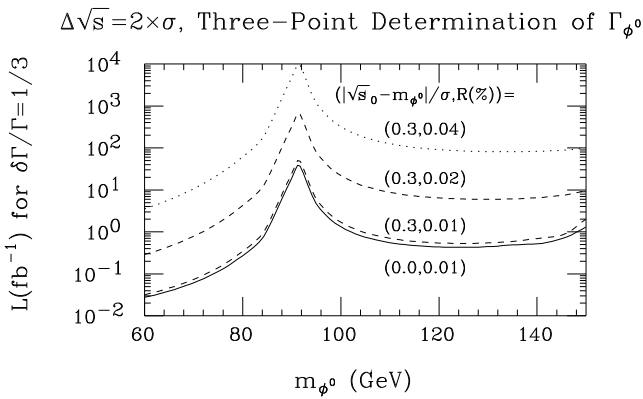

Figure 2: The luminosity required for a $\delta\Gamma/\Gamma = 1/3$ measurement of the $\phi^0$ width vs. $m_{\phi^0}$ for various choices of $(|\sqrt{s} - m_{\phi^0}|/\sigma, R)$.

In addition, the event rate in a given channel measures $\Gamma(\phi^0 \to \mu^+\mu^-) \times BR(\phi^0 \to X)$. Then, using the branching fractions (most probably already measured in $Z\phi^0$ associated production), the $\phi^0 \to \mu\mu$ partial width can be determined, providing an important test of the Higgs coupling. For $L = 50\,\text{fb}^{-1}$ and $R = 0.01\%$, a better than $\pm 1.5\%$ measurement of the $X = b\overline{b}$ channel rate can be performed for all $m_{\phi^0} \lesssim 150\,\text{GeV}$.[4] In obtaining a direct determination of $\Gamma(\phi^0 \to \mu^+\mu^-)$ we will be limited by the $\sim \pm 7\% - \pm 10\%$ measurement of $BR(\phi^0 \to b\overline{b})$ obtained at the NLC by combining the $Z\phi^0$ inclusive rate with the $Z\phi^0 \to Zb\overline{b}$ partial rate (the uncertainty in the inclusive $Z\phi^0$ measurement dominates the error).

A $\mu^+\mu^-$ collider provides two particularly unique probes of the MSSM Higgs sector. First, the couplings of the $h^0$ deviate sufficiently from exact SM Higgs couplings that it may well be distinguishable from the $\phi^0$ by measurements of $\Gamma_h$ and $\Gamma(h \to \mu^+\mu^-)$ at a $\mu^+\mu^-$ collider, using the $s$-channel resonance process (here we use the notation $h$ for a generic Higgs boson). For instance, in the $b\overline{b}$ channel $\Gamma_h$ and $\Gamma(h \to \mu^+\mu^-) \times BR(h \to b\overline{b})$ can both be measured with good accuracy. The deviations for these quantities from SM-Higgs expectations exceed 20% (10%) for $m_{A^0} \lesssim 500\,\text{GeV}$ (700 GeV) for all but small $\tan\beta$ values.[8] Unless $m_h \sim m_Z$, $L = 50\,\text{fb}^{-1}$ of luminosity will yield a better than $\pm 5\%$ determination of $\Gamma_h$, and a better than $\pm 1\%$ determination of $\Gamma(h \to \mu^+\mu^-) \times BR(h \to b\overline{b})$. However, our ability to predict $BR(h \to b\overline{b})$ and $\Gamma_h$ is limited by uncertainty in $m_b$, an uncertainty of order $\pm 5\%$ in $m_b$ leading to a $\pm 3\%$ ($\sim \pm 10\%$) uncertainty in $BR$ ($\Gamma$). If we can keep systematic and statistical errors below $\sim 10\%$, these quantities will probe the $h^0$ vs. $\phi^0$ differences for $m_{A^0}$ values as large as $400-500\,\text{GeV}$.

The second dramatic advantage of a $\mu^+\mu^-$ collider in MSSM Higgs physics is the ability to study the non-SM-like Higgs bosons, e.g. for $m_{A^0} \gtrsim 2m_Z$ the $H^0, A^0$. An $e^+e^-$ collider can only study these states via $Z^\star \to A^0 H^0$ production, which could easily be kinematically disallowed since GUT scenarios typically have $m_{A^0} \sim m_{H^0} \gtrsim 200\text{--}250\,\text{GeV}$. In $s$-channel production the $H^0, A^0$ can be even more easily observable than a SM-like Higgs if $\tan\beta$ is not near 1. This is because the partial widths $\Gamma(H^0, A^0 \to \mu^+\mu^-)$ grow rapidly with increasing $\tan\beta$, implying that $\overline{\sigma}_{H^0, A^0}$ will become strongly enhanced relative to SM-like values. $BR(H^0, A^0 \to b\overline{b})$ is also enhanced at large $\tan\beta$, implying an increasingly large rate in the $b\overline{b}$ final state. Thus, we concentrate here on the $b\overline{b}$ final states of $H^0, A^0$ although the modes $H^0, A^0 \to t\overline{t}$, $H^0 \to h^0 h^0$, $A^0 A^0$ and $A^0 \to Z h^0$ can also be useful.

Despite the enhanced $b\overline{b}$ partial widths, the suppressed (absent) coupling of the $H^0$ ($A^0$) to $WW$ and $ZZ$ means that the $H^0$ and $A^0$ remain relatively narrow at high mass, with widths $\Gamma_{H^0}, \Gamma_{A^0} \sim 0.1$ to 3 GeV; but because $m_{A^0} \sim m_{H^0}$ the $H^0$ and $A^0$ resonance peaks can overlap substantially. Since the individual resonance peaks have width comparable to or broader than the expected $\sqrt{s}$ resolution for $R = 0.06\%$ and $\sqrt{s} \gtrsim 200\,\text{GeV}$, determination of the resonance peak shape would be pos-sible by scanning in $\sqrt{s}$; the $H^0$ and $A^0$ widths could be extracted provided that the signal rates are sufficiently high. The results of a fine scan can be combined to get a coarse scan appropriate for broader widths.

The cross section for $\mu^+\mu^- \to A^0 \to b\bar{b}$ production with $\tan\beta = 2, 5$ and 20 (including an approximate cut and $b$-tagging efficiency of 50%) is shown versus $m_{A^0}$ in Fig. 3 for beam resolution $R = 0.01\%$. Overlapping events from the tail of the $H^0$ resonance are automatically included. Also shown is the significance of the $b\bar{b}$ signal for delivered luminosity $L = 0.1$ fb$^{-1}$ at $\sqrt{s} = m_{A^0}$. Discovery of the $A^0$ and $H^0$ will require an energy scan if $Z^\star \to H^0 + A^0$ is kinematically forbidden; a luminosity of 20 fb$^{-1}$ would allow a scan over 200 GeV at intervals of 1 GeV with $L = 0.1$ fb$^{-1}$ per point. The $b\bar{b}$ mode would yield at least a $10\sigma$ signal at $\sqrt{s} = m_{A^0}$ for $\tan\beta \gtrsim 2$ for $m_{A^0} \lesssim 2m_t$ and at least a $5\sigma$ signal for $\tan\beta \gtrsim 5$ for all $m_{A^0} \lesssim 500$ GeV. The resulting statistical significance for $R = 0.06\%$ is only noticeably worse (by a factor of 2) in the $\tan\beta = 2$ case.[4] For $m_{A^0} \gtrsim m_Z$ ($m_{A^0} \lesssim m_Z$), the $H^0$ ($h^0$) has very similar couplings to those of the $A^0$ and would also be observable in the $b\bar{b}$ mode for $\tan\beta \gtrsim 5$. For $\tan\beta \sim 2$, $BR(H^0 \to b\bar{b})$ is smaller than in the case of the $A^0$ due to the presence of the $H^0 \to h^0 h^0$ decay mode. For such $\tan\beta$ values, detection would be easier in the $h^0 h^0$ final state. Overall, discovery of both the $H^0$ and $A^0$ MSSM Higgs bosons (either separately or as overlapping resonances) would be possible over a large part of the $m_{A^0} \gtrsim m_Z$ MSSM parameter space.

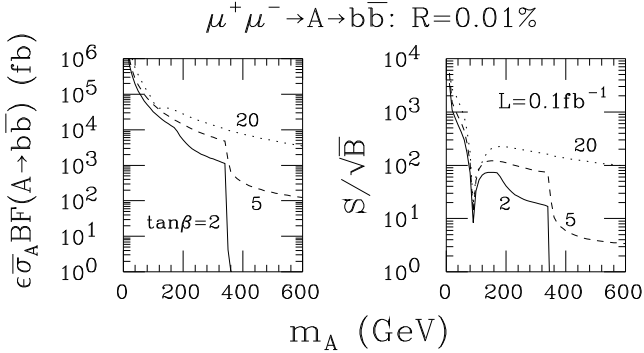

Figure 3: (a) $\epsilon\bar{\sigma}_{A^0} BR(A^0 \to b\bar{b})$, for $s$-channel production of the MSSM Higgs boson $A^0$ versus $\sqrt{s} = m_{A^0}$, for $\tan\beta = 2, 5$ and 20, beam resolution $R = 0.01\%$ and channel isolation efficiency $\epsilon = 0.5$; and (b) corresponding statistical significance of the $A^0 \to b\bar{b}$ signal for $L = 0.1$ fb$^{-1}$ delivered at $\sqrt{s} = m_{A^0}$.

The results we have quoted above do not include the spreading out of the Gaussian luminosity peak due to photon bremsstrahlung. For Higgs bosons with $\Gamma_h \ll \sigma$, the cross section at $\sqrt{s} = m_h$ decreases proportionally to the decrease in the peak luminosity, which in turn depends upon the resolution $R$ and $\sqrt{s}$. Roughly, for any Higgs boson with width much smaller than the energy resolution $\sigma$, a factor of $\lesssim 2$ larger luminosity would be required than given in the numerical results quoted above for any given measurement. For a Higgs boson with width much larger than $\sigma$, the increase in luminosity required for a given measurement would be much less.

Clearly, $s$-channel Higgs production presents exciting possibilities. The techniques discussed are generally applicable to searches for any Higgs boson or other scalar particle that couples to $\mu^+\mu^-$. If any narrow-width Higgs or scalar particle is observed at either the LHC or NLC, a $\mu^+\mu^-$ collider of appropriate energy would become a priority simply on the basis of its promise as a Higgs/scalar factory.

## 2 SUSY

An exciting possibility[2] is that the NMC could be a SUSY factory, producing squark pairs, slepton pairs, chargino pairs, associated neutralinos, associated $H + A$ Higgs, and gluinos from squark decay if kinematically allowed. If the SUSY mass scale is $M_{\rm SUSY} \sim 1$ TeV, many sparticles could be beyond the reach of the NLC. The LHC can produce them, but disentangling the SUSY spectrum and measuring the sparticle masses will be a real challenge at a hadron collider, due to the complex nature of the sparticle cascade decays and the presence of QCD backgrounds. The measurement of the sparticle masses is important since they are a window to GUT scale physics.

The cross sections for squarks (of one flavor in the approximation of $L, R$ degeneracy), charginos, top and three generations of singlet quarks (from an $E_6$ GUT model, for example) are:[2] $\sigma_{\tilde{u}_{L,R}} = 4\beta^3$ fb, $\sigma_{\tilde{d}_{L,R}} = 1\beta^3$ fb, $\sigma_{\chi^\pm} = 6\beta$ fb $\sigma_t = 8$ fb, and $\sigma_{Q_{E_6}} = 6\beta$ fb, leading to 250, 60, 500, 800, and 600 events, respectively, for sparticle masses of 1 TeV, assuming $\sqrt{s} = 4$ TeV and an integrated luminosity of 100 fb$^{-1}$. The production of heavy SUSY particles will give spherical events near threshold characterized by multijets, missing energy (associated with the LSP), and leptons. There should be no problem with backgrounds from SM processes.

A supergravity model with $\tan\beta = 5$, universal scalar mass $m_0 = 1000$ GeV and gaugino mass $m_{1/2} = 150$ GeV provides an illustration of a heavy sparticle spectrum, as follows (GeV units): $\tilde{u}$ : 1000, $\tilde{g}$ : 500, $\tilde{\ell}$ : 1000 $\chi_4^0, \chi_3^0, \chi_2^\pm$ : 350, $\chi_2^0, \chi_1^\pm$ : 130, and $\chi_1^0$ (LSP) : 60. Consider $\tilde{u}\bar{\tilde{u}}$ production at the NMC. The dominant cascade chain for the decays is $\tilde{u}\bar{\tilde{u}} \to (\tilde{g}u)(\tilde{g}\bar{u})$, followed by $\tilde{g} \to \chi_1^\pm q\bar{q}$, with $\chi_1^\pm \to \chi_1^0 \ell\nu, \chi_1^0 q\bar{q}$. The dominant branching fractions of the $\tilde{u}\bar{\tilde{u}}$ final state are

$$\begin{aligned} 10 \text{ jets} + \not{p}_T && 10\% & \\ 8 \text{ jets} + 1\ell + \not{p}_T && 10\% & \quad (1) \\ 6 \text{ jets} + 2\ell + \not{p}_T && 2\% & \end{aligned}$$

Of the two lepton events, one half will be like-sign dileptons ($\ell^+\ell^+, \ell^-\ell^-$). The environment of a $\mu^+\mu^-$ collider may be better suited than the LHC to the study of the many topologies of sparticle events.

Returning to the SUSY Higgs sector, we simply emphasize the fact that $Z^* \to H^0 A^0, H^+ H^-$ will allow $H^0, A^0, H^\pm$ discovery up to $m_{H^0} \sim m_{A^0} \sim m_{H^\pm}$ values somewhat below $\sqrt{s}/2 \sim 2$ TeV. While GUT scenarios prefer $H^0, A^0, H^\pm$ masses above 200 to 250 GeV, such that $H^0 A^0$ and $H^+ H^-$ pair production are beyond the kinematical reach of a 400 to 500 GeV collider, even the most extreme GUT scenarios do not yield Higgs masses beyond 2 TeV. Thus, a 4 TeV $\mu^+\mu^-$ collider is guaranteed to find all the SUSY Higgs bosons.

## 3 Strong $W_L W_L$ Scattering

If a SM-like Higgs boson with $m \leq \mathcal{O}(800\text{ GeV})$ does not exist, then the interactions of longitudinally polarized weak bosons $W_L, Z_L$ became strong. This means that new physics must be present at the TeV energy scale. The high reach in energy of the NMC is of particular interest for study of a strongly interacting electroweak sector (SEWS) at a $\mu^+\mu^-$ collider via $WW$ fusion. The SEWS signals depend on the model for $W_L^+ W_L^-$ scattering. An estimate of the size of these signals can be obtained from the SM by taking the difference of the cross section for a heavy Higgs boson ($m_{\phi^0} = 1$ TeV) and that for a massless Higgs boson: $\Delta\sigma_{\text{SEWS}} = \sigma(m_{\phi^0} = 1\text{ TeV}) - \sigma(m_{\phi^0} = 0)$. The subtraction of the $m_{\phi^0} = 0$ result removes the contributions due to scattering of transversely polarized $W$-bosons. The difference $\Delta\sigma_{\text{SEWS}}$ grows rapidly with energy. At $\sqrt{s} = 1.5$ TeV (for the NLC) $\Delta\sigma_{SEWS}(W^+W^-), \Delta\sigma_{SEWS}(ZZ) = 8$ fb, 6 fb, while at $\sqrt{s} = 4$ TeV (for the NMC) we find 80 fb, 50 fb, respectively. The NMC signals are nearly 10 times larger than the NLC signals.

## 4 Conclusions

There are many other new physics possibilities for which the NMC would be a powerful probe, including

- extra neutral gauge bosons (the NMC could be a $Z'$ factory, with its decays giving Higgses and $W^+W^-$ along with particle and sparticle pairs)
- right-handed weak bosons (the present limit on the $W_R$ of many left-right symmetric models is $M_{W_R} \gtrsim 1.5$ TeV)
- vector-like quarks and leptons (such as those present in $E_6$ models)
- horizontal gauge bosons $X$ (whose presence may be detected as an interference between $t$-channel $X$ exchange and $s$-channel $\gamma, Z$ exchanges; present limits are $M_X \gtrsim 1$ TeV)
- leptoquarks
- compositeness

The list goes on with other exotica. The potentially very large center of mass energy and relative freedom from large backgrounds imply that muon colliders would be very exciting machines for detecting all types of new physics.

## 5 Acknowledgements

I am grateful for the contributions of my collaborators, V. Barger, E. Berger, and T. Han. This work was supported in part by the U.S. Department of Energy.